\documentclass[12pt]{iopart}\usepackage{iopams}  

\usepackage{graphicx}

\usepackage{amssymb,amsfonts}       
\usepackage{mathrsfs}

\usepackage{amsbsy}

\begin{document}
\def\Gcal{\mathcal{G}}
\def\g{\ss}
\def\Po{\wp}
\def\Pob{{\pmb{\Po}}}
\def\Polb{{\pmb{\Po}}}
\def\Polvac{\Po_0 }
\def\Polbvac{\Pob_0 }
\def\Nssub{{\mbox{\tiny${N}$}}}
\def\EM{{\mbox{rep}}}
\def\DR{{\mbox{\tiny{DR}}}}
\def\vel{v}

\def\empt{{\mbox{\tiny${\emptyset}$}}}
\def\Gc{\Gcal}
\def\gsub{g}
\def\kc{{\sf k}} 
\def\ev{\epsilon}
\def\pd{\partial}
\def\vphi{\varphi}
\def\psiR{\widetilde{\psi}}
\def\psiL{\widetilde{\psi}^{{\rm vir}}}
\def\PsimL{{\widetilde{ {\mit \Psi}}}^{{\rm vir}}}
\def\a{\alpha}
\def\uav{\bar{u}}
\def\D{\Delta}
\def\th{\theta}
\def\r{{\mbox{\tiny${R}$}}}
\def\re{{\mbox{\tiny${R}$}}}
\def\Lw{L_{\varphi}}
\def\Efb{{\bf E}}
\def\Bfb{{\bf B}}
\def\W{{\mit \Omega}}
\def\arm{{\rm a}}
\def\brm{{\rm b}}
\def\lf{\left}
\def\rt{\right}
\def\hquad{ \ \ } 
\def\Taum{{\mit \Gamma}}

\title{Gravity between Internally Electrodynamic Particles}

\author{J. X. Zheng-Johansson$^1$ and P-I. Johansson$^2$ 
} 
\date{May, 2007; Feb., 2010}

\address{
1. Institute of Fundamental Physics Research, 611 93 Nyk\"oping, Sweden }
\address{
2. Uppsala University, Studsvik, 611 82 Nyk\"oping, Sweden}

\begin{abstract}
We present a first-principles' prediction that two charged particles  of masses $M_1$ and $M_2$ separated $R$ apart in a dielectric vacuum act on each other always an  attractive force in addition to other known forces in between. This component attractive force on one charge results as the Lorentz force in the radiation depolarization- and  magnetic- fields of the other charge,  being an attractive radiation force, and is in addition to the ordinary  repulsive radiation force.  The   exact solution for the attractive radiation force is $F_\gsub=\Gc M_1M_2/R^2$,  an identical formula to Newton's law of gravitation.  $\Gc={\cal X}_{0^*}e^4 /4\pi \epsilon_0^2\hbar^2$$\rho_l$ is identifiable with  Newton's gravitational constant, ${\cal X}_{0^*}$ being  the susceptibility and $\rho_l$ the linear mass density of the vacuum, and  the remaining  fundamental constants of the usual meaning. The $F_\gsub$ force  is conveyed by a transverse vacuuonic dipole-moment wave traveling at the velocity of light and can penetrate matter freely. In all of respects, the $F_\gsub$ force represents a viable cause of Newton's universal gravity.\footnote{This edition mainly contains update of certain terminology in current use. A more comprehensive treatment of the gravity of IED particle  is in preparation. JXZJ, Feb, 2010.}

\end{abstract}

%



\maketitle

\def\citeUnif{3}

Since  Sir. Isaac Newton discovered in 1986\cite{Newton:1686} gravity described by an inverse square law, the cause of this feeble but ubiquitous fundamental force between all bodies  has remained a subject of many studies \cite{Whittaker:1951:I28}. 
 The cause of gravity has remained as an open question to date.
 One of us (JXZJ) recently,   based on overall experimental  information, developed an internally electrodynamic (IED) particle model  [\citeUnif a--m] (earlier termed a basic-particle formation scheme) along with a model construction of the vacuum.
 The first-principles' solutions [\citeUnif a--m] predict a range of particle properties and their  relations, including an attractive radiation force, directly comparable with  observations and the established basic laws of physics. In the present paper we elaborate on the solution in detail; we shall show that the  solution for the DR force predicts an inverse square formula identical to Newton's law of gravitation; the proportionality constant is expressed solely by fundamental constants or constants of the medium. As a major extension to the ealier conference paper[\citeUnif f], in this paper we also present a quantitative determination of the electric susceptibility of vacuum.

 \def\sang{{\mathscr{Q}}}
\def\D{\Delta}
\def\Ab{{\bf A}}
\def\Rxb{{\bf R}}
\def\Efb{{\bf E}}
\def\Bfb{{\bf B}}
\def\Bb{{\bf B}}
\def\Fb{{\bf F}}
\def\Rb{{\bf R}}
\def\Fepionii{F_{\pol_{12}}}
\def\Fepiioni{F_{\pol_{21}}}
\def\Fbepionii{\Fb_{\pol_{12}}}
\def\Fbepiioni{\Fb_{\pol_{21}}}
\def\Rbi{\Rb_1}
\def\Rbii{\Rb_2}
\def\Xim{{\cal X}}
\def\ke{{\sf k}}
\def\kvac{\ke_0}
\def\thuni{{\hat \theta}}
\def\Zuni{{\hat Z}}
\def\th{\theta}
\def\phiuni{{\hat \phi}}
\def\W{{\mit \Omega}}
\def\Rbi{\Rb_1}
\def\Rbii{\Rb_2}
\def\Kb{{\bf K}}
\def\Eb{\Efb}
\def\th{\theta}
\def\phiuni{{\hat \phi}}
\def\Rx{R}
\def\qi{q_1}
\def\qii{q_2}
\def\Wi{\W_1}
\def\Wii{\W_2}
\def\Tm{T}
\def\Ki{K_1}
\def\hquad{ \ \ } 
\def\pd{\partial}
\def\Ef{E } 
\def\a{\alpha}
\def\Amp{A}
\def\Ampi{A_1}
\def\Eng{{\cal E}}
\def\phiv{\varphi}
\def\Taum{{\mit \Gamma}}
\def\Lw{L_{\rm \Ac}}
\def\Ac{ \varphi}
\def\l{l}
\def\lf{\left}
\def\rt{\right}
\def\M{M}
\def\Mi{M_1}
\def\Mii{M_2}
\def\Mcal{{\mathfrak{M}}}
\def\lsub{l}
\def\Efbpoli{\Efb_{\pol 1} }
\def\Efbpolii{\Efb_{\pol 2} }
\def\pol{{\rm p}}
\def\vbpolii{\vb_{p2}}
\def\vb{{\bf v}}
\def\De{\Delta}
\def\Fbmpionii{\Fb_{g_{12}}}
\def\Fbmpiioni{\Fb_{g_{21}}}
\def\Xuni{{\hat X}}
\def\Kii{K_2}
\def\v{{\rm v}}

\begin{figure}[b]
\vspace{0.2cm}
\centering
\includegraphics[width=0.65\textwidth]{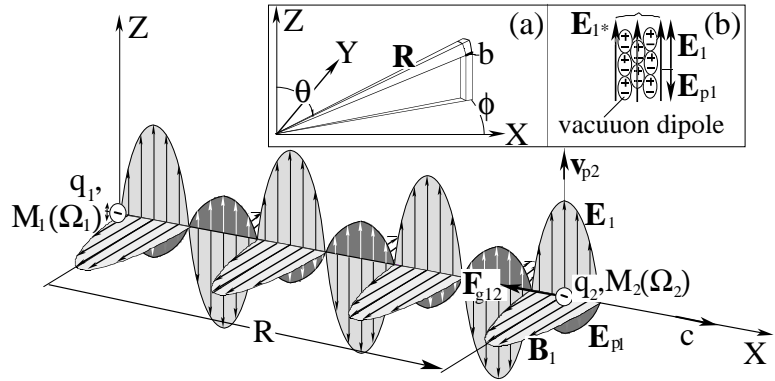}
\caption{
Oscillatory charge $q_1$ produces at $R$ a radiation electric and magnetic fields $\Efb_1(R,T)$, $\Bfb_1(R,T)$, and in a dielectric vacuum also a radiation depolarization field $\Efb_{\pol 1}(R,T)$. Consequently charge $q_2 $  of dynamical mass $M_2(\W_2)$ at  $R$ is acted the component Coulomb force  $q_2\Efb_{\pol 1}$, driven into motion at a component velocity $\vb_{p2}$, each along $Z$-axis. $q_2$ is in turn acted by an attractive radiation Lorentz force $\Fb_{g12}=q\vb_{p2}\times \Bfb_1$ in $-X$-direction;  $\Fb_{g12}$ points  toward $M_1$ and is an attraction always. Similarly (not shown in the plot),  charge $q_1$ is acted by an attractive  radiation  Lorentz force $\Fb_{m21}$ due to the fields of $q_2$. Inset (a) shows a radial wave path ${\bf R}$ making an angle $\th$ at the charge oscillation $Z$-axis;  (b) shows $\Eb_{p1}$ in relation to the radiation electric field due to $q_1$, $\Eb_{1^*}$  measured in free space, and  $\Eb_1$ in vacuum. 
}
\label{fig-grv1} \label{fig-grv} 
\end{figure}

Consider two IED model particles, $i=1,2$ located at $\Rb_1=0$ and $\Rb_2$ along $X$-axis (FIG. \ref{fig-grv1});  $R=|\Rxb|= |\Rbii-\Rbi|$. The  IED particles  [\citeUnif a, e,g, i-j] are each composed of  an oscillatory charge $q_i$ of zero rest-mass and the resulting electromagnetic waves. The charge is at its creation in the vacuum initially  endowed with a kinetic energy $\Eng_{qi}$, and in the potential well due to a dielectric vacuum polarized by the charge's own  field,  executes an oscillation. This is  about fixed site  $\Rb_i$ assuming $\Eng_{qi}$ below a threshold value, with an angular frequency  $\W_{qi}$ characteristic of  $\Eng_{qi}$ and the charge-vacuum interaction. Given the vacuum has  only a single energy level as in observation, $\Eng_{qi}$ can not be dissipated  except in a pair annihilation. The radiation displacement (out of a total that is predominately non-radiative), assuming  small, will be effectively sinusoidal,  $ u_{qi}(T)= A_{qi} \sin (\W_{qi} T) $, with $A_{qi}$  the amplitude. $u_{qi}(T)$ has a specified orientation during a finite  time, but will vary in a random fashion over long time under the influence of environmental fields. Except in the context of Eq. (\ref{eq-GrvX1}) later, we shall assume $u_{qi}(T)$ is along the $Z$- axis for a duration being finite but short compared to the time of measurement of say their interaction forces.

Either charge, say 1,  will  owing to its oscillation generate  electromagnetic waves, being monochromatic for the oscillation about fixed site $\Rxb_1$. In three dimensions the wave is propagated in radial directions, along  radial path ${\bf R}(R, \th, \phi)$ at angles  $\th$ and $\phi$ at the $Z$- and 
$X$-axes, see  inset (a) of FIG. \ref{fig-grv1}. 
For the classical radiation in question here, the component  radiation electric  field $\Eb_{1} $ and magnetic field $\Bb_{1}$ 
are given by the solutions to  Maxwell's equations,
            $\Eb_{1}(\Rx,\th,\Tm)=\frac{E_{10} j({\bf l},T)}{R} \sin \theta \hat{\theta}$.
            %
Here, 
$$\displaylines{
\refstepcounter{equation}\label{eq-E10}
\hfill E_{10}=\frac{q_1\W_1^2 A_{q1}}{4 \pi  \epsilon_0 c^2}, \hfill (\ref{eq-E10})
\cr\refstepcounter{equation}\label{eq-jRT}\label{eq-f1}
\hfill     j_1({\bf R},T)= \sin (\Wi \Tm -{\bf\Ki} \cdot {\bf R}+\alpha_1), 
\hfill (\ref{eq-f1})
}$$
  $\epsilon_0$ is the permittivity of the vacuum and  $c$ the speed of light; $\Wi =\W_{qi}$;
$\Wi $ and $K_1=\Wi/c$ are the normal mode angular frequency  and wavevector,  and $\a_1$ is the initial phase. Along the $X$-axis on which  charge 2  lies,   $\th=\pi/2$ and  $\phi=0$; hence $\sin  \th= 1 $,   
$\Kb_1  \cdot {\bf R}
= K_1 R $, and  
$$\displaylines{
\refstepcounter{equation}\label{eq-ERT}\label{eq-Efbi} 
\hfill
\Eb_{1}(\Rx,\Tm)=\frac{ E_{10}    j_1(R,T) }{ \Rx }    {\hat Z},
\quad
 \Bfb_1 (\Rx,\Tm) = -\frac{\Ef_1 }{c}  {\hat Y}.  \hfill
(\ref{eq-ERT})
  }$$

In the dielectric vacuum  $E_1(0,\Tm)$, $\equiv E_{q1}(\Tm)$, represents  a macroscopic field  at $R=0$. This conjures an applied field $E_{q1^*}(\Tm)$ measured in a space after removal of the vacuum medium, termed our {\it free space}. 
As a generalization [\citeUnif a,c] of the dielectrics of ordinary materials\cite{Bottcher}  to a nonpolar dielectric  vacuum, relevant to here we can write down these:  $ E_{q1^*}$ produces in the vacuum a polarization $\Polb_{q1}=\epsilon_{\empt}(\Efb_{q1^*}-\Efb_{q1})$ with $E_{q1}(T)= E_{q1^*}$ $/\ke_{0^*} $, and thus a depolarization field[\citeUnif a, c]: 
$$\Efb_{p.q1}(T)=-\frac{\Polb_{q1}(T)}{\epsilon_{\empt}}
= -\Xim_{0^*} \Efb_{q1}(T).$$
Here $\epsilon_{\empt}$ is the permittivity, $\ke_{0^*}$ the dielectric constant and $\Xim_{0^*}=\ke_{0^*}-1 $ the electric susceptibility of the vacuum,  each being measured  with respect to the free space. Along the chain of the polarized vacuuons, their dipole displacements are mutually coupled as with their center-of-mass motions;  so the disturbance $\Efb_{p.q1}(\Tm)$  will be propagated as a radiation depolarization field: 
$$\displaylines{
\refstepcounter{equation}
\label{eq-Ebp1}\label{eq-Efbip}
\hfill 
\Eb_{p1}(\Rx, \Tm)=-  \Xim_{0^*} \Eb_{1}(\Rx, \Tm).
\hfill (\ref{eq-Efbip})}$$
Assuming $E_{1}$ is relatively small and is real, $E_{p1}(R, \Tm)$ is therefore linear and is in phase with $E_{1}(R,\Tm)$, as shown in FIG. \ref{fig-grv1}. The radiation due to charge 2 can be written down similarly.


The wave variables transform to particle variables as follows. The total electromagnetic wave described by the fields $E_i$,$B_i$ above of frequency $\Omega_i$, has according to M. Planck an energy   $\Eng_i=\hbar \W_i$. The wave together with its generating charge, $i$, makes up 
our IED particle $i$. Therefore the wave is  here the internal components of particle $i$. Consequently its energy $\Eng_i$ represents  the total rest  energy of the particle (with its source charge oscillating about fixed site); and this gives  in turn the particle's rest mass $M_i$ given by $\Eng_i=M_ic^2 $ from direct solution for  the 
IED process [\citeUnif a, g,j].
The two  equations of $\Eng_i$ give
$$ 
\refstepcounter{equation}\label{eq-W12}
\W_i^2 =M_i^2 c^4/\hbar^2, \qquad i=1,2. 
          \eqno(\ref{eq-W12})
$$
 $\Eng_i $ is nondissipative as is  $\Eng_{qi}$, and it retains with charge $i$ through wave reflections from surrounding objects. We can  describe the electromagnetic waves alternatively  as a heuristic technique  only here  as elastic waves (see a systematic justification and representation given in [\citeUnif a]), along a chain of coupled vacuuonic oscillators of size $b$ each. The standard solution of classical wave mechanics, [\citeUnif a,g,h], gives the total wave energy integrated over all directions across a total wave train length  $L_\varphi$,   thus  $L_\varphi/b=\frac{c \D T}{b}$ oscillators, in time $\D T$, to be $$\refstepcounter{equation} \label{eq-E1}
 \Eng_i=\frac{1}{8 \pi } c \D T  \rho_\l \W_i^2 A_{qi}^2, \qquad i=1,2.  
\eqno (\ref{eq-E1})
$$ 
The $\Eng_i$ of (\ref{eq-E1}) and the $\Eng_i=M_ic^2$ earlier together yield 
$$   \refstepcounter{equation} \label{eq-GrvX1}
\W_i^2 A_{qi}^2 = 8 \pi M_i c/ \D T   \rho_{\lsub}.  
\eqno(\ref{eq-GrvX1})
$$

Charge 2 is in the $\Efbpoli $ field of charge 1 subject to the Coulomb force $\Fbepionii = \qii \Efbpoli $. Suppose in time ($T-\frac{\D T}{2}, T+\frac{\D T}{2}$) particle 2 is driven from at rest into motion of a component velocity $\vbpolii (\D T)$. This substituted in Newton's second law, $\Fbepionii = M_2\frac{\pd \vb_{p2}}{\pd T}$, gives the equation of motion of particle 2: 
$
\qii \Efbpoli = $ $ \Mii \frac{ \pd \vbpolii }{ \pd \Tm } 
$. (We shall not be concerned with the motion due to $\Eb_1$ until a discussion in the end.)
The integral velocity in time $\De \Tm$ is $\vbpolii (\D T) =\int_0^{ \vbpolii (\D T)} d \vbpolii =  \frac{\De \Tm \qii \Efbpoli  }{\Mii}
$; 
with (\ref{eq-Ebp1}) this becomes
$$\displaylines{
\refstepcounter{equation}
\label{eq-Grv19} 
\hfill
\vbpolii  (\De \Tm) 
 = -\frac{\De \Tm \qii   \Xim_{0^*}  \Efb_1 (R,T)  }{\Mii} 
\hfill (\ref{eq-Grv19})
}$$
In the $\Bfb_1$ field the moving charge 2 is further  acted by a force according to the Lorentz force law: 
$$\displaylines{
\refstepcounter{equation}
  \label{eq-Grv-A4}
\hfill
\Fbmpionii (\Rx,  \Tm)= \qii \vbpolii (\Rx,  \De \Tm) \times  \Bfb_1 (\Rx,  \Tm) 
= -|\qii v_{p2}  B_1| \hat{X}      
\hfill (\ref{eq-Grv-A4})
}$$
$\Fbmpionii$ is an attractive  
radiation force resulting from the radiation depolarization  electric field and magnetic field of charge 1, and
          being   antiparallel with the ordinary repulsive radiation force. 
The direction of $ \Fbmpionii$ is  perpendicular to $\vb_{p2}$ and $\Bb_1$ (the right-hand rule), lying  along the $X$-axis. The second expression of (\ref{eq-Grv-A4})  holds always valid for $\vbpolii \bot \Bfb_1$ here. $\vbpolii \times  \Bfb_1<0$ for $q_2>0$; and $\vbpolii \times \Bfb_1>0$ for $q_2<0$ (as in FIG. \ref{fig-grv1}). That is,  in either case $ \Fbmpionii$ is pointed toward mass $M_1$, and is an attraction.

Now with (\ref{eq-Grv19}) for $\vb_{p2}$ and (\ref{eq-Efbi}) for $\Bfb_1 $ in (\ref{eq-Grv-A4}), we have $\Fbmpionii = - \frac{(\De \Tm) \qii^2   \Xim_{0^*}  \Ef_1^2  }{\Mii c}   \  \Xuni$. Further  with (\ref{eq-Efbi}) for $\Ef_1$, this becomes   
$$\refstepcounter{equation} \label{eq-F12a}
\hfill  
\Fbmpionii (\Rx, \Tm)
= -\frac{(\De \Tm)  \Xim_{0^*}   \qi^2 \qii^2 \Wi^4   A_{q1}^2  }{
(4 \pi)^2     \epsilon_0^2 c^5\Mii  }  \frac{  j_1^2(R,T)      }{\Rx^2 } \hquad   \Xuni     
                   \hfill \eqno (\ref{eq-F12a})
$$
As an elementary consideration here we put $q_1, q_2= +e$ or $-e$, with $e$ the elementary charge. With accordingly $e^4$ for $q_1^2q_2^2 $, and in turn  (\ref{eq-W12})  for $\W_1^2$, and (\ref{eq-GrvX1}) for $\W_{1}^2 A_{q1}^2$  in ({\rm a}), simplifying algebraically, we have 
$$
\refstepcounter{equation}\label{eq-Grv21b}
 \Fbmpionii (\Rx,\Tm) 
= -\frac{2\Gc M_1^3 j^2_1(R,T)}{M_2 R^2}\hquad \Xuni,   
                         \eqno(\ref{eq-Grv21b})
$$
where
$$
 \refstepcounter{equation}\label{eq-Gv26}
\Gc = \frac{   \pi \Xim_{0^*}    e^4    }{
\epsilon_0^2 h^2   \rho_{\lsub}  }. \eqno(\ref{eq-Gv26})
$$
Or with $c=1/\sqrt{\mu_0\epsilon_0 }$, (\ref{eq-Gv26}) is written  as
$\Gc = \frac{\pi \Xim_{0^*} \mu_0^2   e^4  c^4   }{
 h^2   \rho_{\lsub}  }. 
$ 
We can likewise regard charge  2 as the source and get similarly an always attractive radiation Lorentz force acting on particle 1 due to  2:
$$\displaylines{
\refstepcounter{equation}
 \label{eq-Grv24}
\hfill
\Fbmpiioni (\Rx,\Tm) = - \Fbmpionii (\Rx,\Tm)
= \frac{2\Gc M_2^3 j_2^2(R,T)}{M_1 R^2}
 \hquad \Xuni,  \hfill (\ref{eq-Grv24})
}$$
with $j_2(R,T)=\sin (\Wii \Tm +\Kii \Rx +\a_2)  $ and 
$\a_2$ the initial phase of the EM wave due to charge 2.

As  expressed in (\ref{eq-Grv-A4}) and (\ref{eq-Grv24}), 
for $q_i$ and $q_j$ ($i,j=1,2$) having any signs, and moved to any locations  relative to each other, {\it the attractive radiation Lorentz force  }  {\it acting on charge i due to (the radiation fields of) the charge j}, $\Fb_{gij}$, {\it is always attractive}. An always attractive force $F_{g12}$  follows also from the fact that it has a field $\Eb_{p1}$ always antiparallel with $\Efb_1$ that together with the common $\Bfb_1$ produces a repulsive ordinary  radiation force.

The dynamical variables  $\Fbmpionii$ and $\Fbmpiioni$ are mutually uncorrelated and statistical, as  $\a_1$ and $\a_2$ are. Their total action as sampled at one time is therefore the {\it geometric mean} 
$$\displaylines{
\refstepcounter{equation}
 \label{eq-grv30}
\hfill 
{\widetilde F}_\gsub (\Rx,\Tm) 
= \sqrt{ |\Fbmpionii (\Rx,\Tm) | |\Fbmpiioni (\Rx,\Tm) | }  
=\frac{ 2\Gc M_1 M_2 j_1^2j_2^2 }{\Rx^2 }.  
\hfill (\ref{eq-grv30})}$$
We are furthermore interested in the average of ${\widetilde F}_\gsub  $ over a macroscopic  time comparable with experimental measurement, or practically over the wave periods $\Taum_1$ and $ \Taum_2$ which suffices for the periodic processes here. This is: $F_\gsub=(1/\Taum_1\Taum_2) \int_0^{\Taum_1} d T\int^{\Taum_2}_0
 {\widetilde F}_\gsub (R,T)d T$; or,  
$F_\gsub =  \frac{ 2\Gc M_1 M_2 I}{\Rx^2 }   $ 
where 
$ I 
=
(\frac{1}{\Taum_1}    \int_0^{\Taum_1} 
 j_1(R,T) 
  \ d \Tm )^2
=\frac{1 }{2}$.
Hence
$$\displaylines{\refstepcounter{equation} \label{eq-Gv25}
\hfill
F_\gsub = \frac{  \Gc\Mi \Mii  }{\Rx^2 }. 
\hfill (\ref{eq-Gv25})
}$$
$F_\gsub$ is {\it always an attractive force} given  that $\Fbmpionii $,  $ \Fbmpiioni$ and  ${\widetilde F}_\gsub $ are attractive, and it is described by an inverse square formula  (\ref{eq-Gv25}) identical to Newton's law of gravitation. The precoefficient  $\Gc$ is expressed in (\ref{eq-Gv26}) by solely  fundamental constants, $e$, $\epsilon_0$, and $\hbar$,  and the constants $\Xim_{0^*}$ and $ \rho_{\lsub}  $ characteristic of the vacuum. $\Gc$ follows therefore to be a constant, in direct resemblance to the universal gravitational constant $G$. Also, it can be readily checked that $\Gc$ given by (\ref{eq-Gv26}) has  a dimension m$^3$ kg$^{-1}$ s$^{-2}$ correctly as required by  formula (\ref{eq-Gv25}), or, it is the same as $G$.

In all, the 
attractive radiation force $F_g$ given in (\ref{eq-Gv25}) is a direct analogue of Newton's gravitational force, in its functional form,  its sign,  and the constancy  of its $\Gc$. 
         It is therefore plausible that the underlying physics presents a viable  cause of Newton's universal gravity. Subsequently   $\Efb_{pi}(\Rx,  \Tm) $ and   $\Bb_i(\Rx,  \Tm)$, making up a dipole wave---a periodic wave motion of the  vacuuon dipole moments, represents a gravitational wave. Following (\ref{eq-ERT})--(\ref{eq-Efbip}), this is a transverse wave and has a wave velocity $c_g =\frac{d \W_i}{d K_i}=c$, with $c$ the velocity of light as before; $\frac{d \W_i}{d K_i}=\frac{\W_i}{ K_i}$ in the linear dispersion region of $\W_i$.
Put  $\Gc\equiv G$, with $G$ the gravitational constant; thus (\ref{eq-Gv26}) writes 
$$G= \frac{\pi \Xim_{0^*}    e^4    }{
 \epsilon_0^2  h^2   \rho_{\lsub}  }; 
$$ 
accordingly
$$
F_\gsub = \frac{  G\Mi \Mii  }{\Rx^2 }.
                           \eqno(\ref{eq-Gv25})'
$$
With the known values for $G$ ($=6.673\times10^{-11}$ m$^3$kg$^{-1}$ s$^{-2}$),  $e$, $\epsilon_0$ and $\hbar$,  
we get an estimate  
$$\displaylines{
\refstepcounter{equation}\label{eq-ratio}
\hfill
\rho_\l /\Xim_{0^*}=\pi e^4/( \epsilon_0^2 h^2 G) 
=9.00(8) \times 10^{23} 
\hquad {\rm kg/m}.       
\hfill (\ref{eq-ratio})
}$$

Besides the two earlier evaluations, the total wave energy in time $\Delta T=\frac{L_\vphi}{c}$ is more immediately given by the standard result computed from radiation electromagnetic  fields, $\Eng_{i}=
\frac{1}{4\pi \Taum_i} 
\int_0^{\Taum_i}
\int_0^{4\pi}
\epsilon_0 E_i^2(R,\th,T) 
R^2 d \sang d T 
= \frac{L_\phi e^2 \W_i^4 A_{qi}^2   }{96\pi^2 \epsilon_0 c^4}
 $. This combined with (\ref{eq-E1}) yields: $\rho_{l i} =\frac{\pi \mu_0 e^2 c^2M_i^2}{3h^2  }$. A vacuum region between two masses $M_i$ and  $M_j$ is described by the mean value 
$$\displaylines{
\refstepcounter{equation} \label{eq-rhol}
\hfill 
\rho_l=\sqrt{\rho_{l i}\rho_{l j}} =\frac{\pi \mu_0 e^2 c^2M_iM_j}{3h^2}, \qquad i,j=1,2.
\hfill(\ref{eq-rhol})
}$$
(\ref{eq-rhol}) and (\ref{eq-ratio}) enable us to determine $\Xim_{0^*}$ directly:
$$\displaylines{
\refstepcounter{equation}
 \label{eq-Chi}
\hfill
\Xim_{0^*}=\frac{\epsilon_0 G M_iM_j}{3e^2}, \qquad i,j=1,2.
\hfill (\ref{eq-Chi})
}$$
$\Xim_{0^*} $ of (\ref{eq-Chi}) and $\rho_l$ of (\ref{eq-rhol}) are seen to both increase with $M_iM_j$, while their ratio $\rho_l/\Xim_{0^*}$ as of (\ref{eq-ratio}) is a constant.

Any material body is representable as a matrix of charges of alternating sings in general, of  equal quantities in a neutral body. The total  attractive radiation force between two such bodies is thus straightforwardly given by the vector sum of those between two charges from the  respective bodies, over all charges in each body, provided the force penetrates all of material objects on its way. That the DR force has  indeed such a superpenetration power 
is elucidated in [3b].

In addition to the $F_g$, two single charges $1,2$ are always also repelled by the usual Lorentz force  due to their fields $E_1,E_2$: $F_{{\rm r}}= \frac{F_g}{\Xim_{0^*}}$. For between an electron mass $M_e $ and proton  mass $ M_p$, for example, (\ref{eq-Chi}) gives
$
\Xim_{0^*}(e,p)=1.17\times 10^{-41}<<1.
$
So $F_{{\rm r}}>>F_g$, giving a net repulsion. However, for two  macroscopic masses, say $M_a=M_b=1$ kg, (\ref{eq-Chi}) gives 
$$\displaylines{
\refstepcounter{equation}\label{eq-Xim1}
\hfill
\Xim_{0^*}(a,b)=7.67\times 10^{15}.
\hfill (\ref{eq-Xim1})
\cr
\mbox{Therefore}\hfill
\cr
\refstepcounter{equation}\label{eq-Fr}
\hfill
F_{{\rm r}}= \frac{F_g}{7.67\times 10^{15}} <<F_g.
\hfill (\ref{eq-Fr})
}$$
Considering also, for their macroscopic sizes, say of radius $r_\alpha=0.1$ m, $\alpha=a,b$, $F_{{\rm r}}$ due to the enormous number, of order $10^{27}$,  of particles in the interim of each body will be effectively shielded off; the surface population is only a fraction $r_\alpha /a \sim 10^{-10}$ of the body (taking $a=5 \times 10^{-11}$ m). In total, the two forces scale as
$
\frac{F_{{\rm r}}    }{F_g} \cdot \frac{a}{r_a}\sim  10^{-26}.
$ 
Consequently, two macroscopic sized (neutral) bodies interact with a net  attraction. 

The gravity theory for IED particle  (earlier editions [\citeUnif f]) is conceived and developed by Dr Zheng-Johansson within a unification project. 
Began in April 2003 (-2006), Prof R. Lundin kindly agreed to have  IRF as a tentative host location for a   VR  funding application   for Dr Zheng-Johansson's unification project, with also a tentative prospect for  an academic position. 
 The research in practice is privately funded by scientist P-I Johansson. We thank  Professor VK Dobrev for having the work presented at QTS 4, Vana, 2005, and the discussions of Professors   H-D Doebner and  G Goldin since our meeting at QTS 4  in connection with  the IED model and solutions compared with quantum mechanical properties.  

\vfill
\eject \vfill
\eject
\newpage


\vfill
\end{document}